\documentclass[twocolumn,showpacs,preprintnumbers,amsmath,amssymb]{revtex4}

\usepackage{graphicx}
\usepackage{dcolumn}
\usepackage{bm}
\usepackage{natbib}

\begin{document}

\preprint{APS/123-QED}

\title{An association between anisotropic plasma heating and instabilities in the solar wind}

\author{J.~C.~Kasper}
\email{jkasper@cfa.harvard.edu}
\affiliation{Harvard-Smithsonian Center for Astrophysics}

\author{B.~A.~Maruca}
\affiliation{Harvard-Smithsonian Center for Astrophysics}

\author{S.~D.~Bale}
\affiliation{Physics Department and Space Sciences Laboratory, University of
  California, Berkeley}

\date{\today}


\begin{abstract}
  We present an analysis of the components of solar wind proton
  temperature perpendicular and parallel to the local magnetic field
  as a function of proximity to plasma instability thresholds.  
  We find that $T_{\perp p}$ is 
  enhanced near the mirror instability threshold 
  and $T_{\parallel p}$ is enhanced 
  near the firehose instability threshold.  
  The increase in $T_{\perp p}$ is consistent 
  with cyclotron-resonant heating, but no similar explanation for 
  hot plasma near the firehose limit is known.  One possible explanation is 
  that the firehose instability acts to convert bulk energy into thermal
  energy in the expanding solar wind, a result with significant 
  implications for magnetized astrophysical plasma in general.
\end{abstract}

\maketitle

{\it Introduction.}---  
Particle velocity distribution functions
in the solar corona and solar wind are anisotropic, 
with separate temperatures $T_\perp$ and $T_\parallel$ 
relative to the magnetic field {\bf B} \citep{eviatar70}.  
Characterizing the processes that create and limit
temperature anisotropy
is important for understanding heating and 
dynamical effects in solar physics
\citep{klimchuck06}
and in astrophysical plasmas in general
\citep{schekochihin09}.
Comprehensive {\it in situ} measurements of the solar wind
by spacecraft are a unique way to investigate
anisotropic plasmas and provide 
observational constraints for more exotic  
astrophysical objects
such as accretion disks \citep{sharma07} and blazars
\citep{roken09}.
The range of anisotropy seen in the solar wind
is a result of competing phenomena: adiabatic 
expansion, heating through the anisotropic 
dissipation of waves,  
instabilities, and Coulomb collisions.  Each of 
these factors has individually been the subject of intensive study;
however, little work has been done on how the effects interact with
and regulate each other.
Our purpose here is to present the 
first observations of these interactions by
looking at heating in the presence of instabilities.  
We begin with a 
review of anisotropy in the solar wind.

If solar wind protons expanded adiabatically, they would 
conserve the first and second CGL invariants so that
$T_{\perp p}\propto B$ and $T_{\parallel p}\propto n_p^2/B^2$ \citep{chew56},
where $p$ denotes protons
and $n_p$ is the number density.  We would then expect
$R_p\equiv T_{\perp p}/T_{\parallel p}\propto B^3/n_p^2$.
The wind would then evolve along a particular
trajectory in $(\beta_{\parallel p},R_p)$-space, 
where $\beta_{\parallel p}=n_pk_BT_{\parallel p}/(B^2/2\mu_0)$
is the ratio of the parallel pressure of 
protons to the magnetic pressure \citep{matteini07}.  
Observations have shown that while
$R_p$ decreases and $\beta_{\parallel p}$ increases with distance 
from the Sun, the slope of the trajectory is 
inconsistent with adiabatic expansion.  Specifically, an additional
source of perpendicular heating must be introduced
\citep{matteini07}.  

Evidence for perpendicular heating is common in regions such as
the ionosphere, the solar wind, and the solar corona, where ions 
may develop $R>20$ \citep{cranmer08}. One 
explanation is 
Alfv\'{e}n-cyclotron dissipation, in which ions enter cyclotron 
resonance with compressive Alfv\'{e}n waves on spatial scales near their 
gyro-radius and are preferentially energized 
perpendicular to {\bf B} \cite{hu99, cranmer03,isenberg07}.  
Ion heating consistent with an Alfv\'{e}n-cyclotron mechanism 
is also directly observed in interplanetary space
\citep{kasper08}.

\begin{figure}
  \includegraphics[scale=.50]{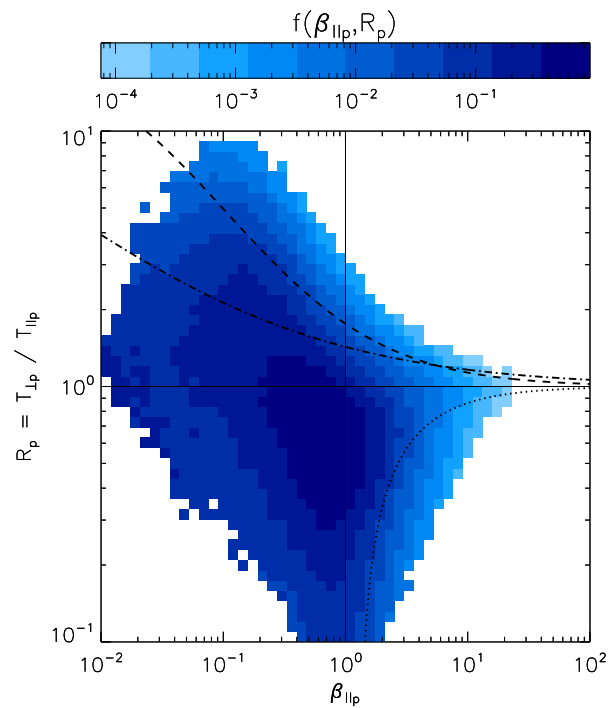}
  \caption{The distribution function $f$ of the solar wind in 
    $(\beta_{\parallel p},R_p)$-space at 1 AU.  
    Lines are theoretical curves of constant
    growth rate for  
    the firehose (dotted), mirror (dashed), and cyclotron (dot-dashed)
    instabilities.  These instabilities bound a stable range of $R_p$ as 
    a function of $\beta_{\parallel p}$.  
    \label{fig1}}
\end{figure}

Instabilities prevent expansion and heating from
driving $R_p$ arbitrarily far from unity.
Fig.~\ref{fig1} is the probability distribution $f$ of the solar wind
in $(\beta_{\parallel p},R_p)$-space as measured by the {\it Wind} 
spacecraft at 1 AU.  These observations
are the focus of this
letter and are described in more detail in the next section.
Given the large anisotropies in the corona and the
effects of adiabatic expansion, one would expect 
$R_p$ to vary over orders of magnitude and be 
as small as $10^{-3}$ \citep{matteini07,bale09}.  Instead we find
$0.1\lesssim R_p\lesssim 5$ and constrained to a narrowing region near 
isotropy as $\beta_{\parallel p}$ increases.  
This narrowing of $f$ with $\beta_{\parallel p}$ is 
attributed to instabilities, driven by $R_p$, 
that generate 
electromagnetic fluctuations, scatter particles in velocity space,
and drive $R_p$ toward unity. 
For $R_p<1$ and $\beta_{\parallel p}\gtrsim 1$ the firehose 
instability can limit anisotropy \citep{kasper02}, while for
$R_p>1$, the mirror and cyclotron instabilities 
are active \citep{gary94,hellinger06}. 
One way to quantify instabilities is to 
calculate the rate of growth of the unstable modes
of the linear Vlasov equation.
The three curves in Fig.~\ref{fig1} are 
contours of constant growth rate $10^{-3}$ times the 
proton cyclotron frequency for 
each instability \citep{hellinger06}.  
The sharp drop in $f$ beyond these curves is 
interpreted as evidence of the instabilities.  Comparing the 
shape of $f$ with these curves,
the mirror instability appears to be more important
than the cyclotron instability at limiting $R_p>1$, even though the
amplitude of the cyclotron instability grows more quickly for 
$\beta_{\parallel p}\le 2$.  This diminished role for the cyclotron 
instability is supported by the recent discovery 
of enhanced magnetic fluctuations in plasma beyond the  
mirror and firehose threshold curves, but not the cyclotron 
threshold \citep{bale09}.
This surprising result, which highlights the 
potential pitfalls of linear
theory, is possibly due to mirror fluctuations being 
more efficient at scattering particles in velocity 
space even if the growth rate is slower \citep{hellinger06}.

Having introduced the heating mechanisms and instabilities 
associated with temperature anisotropy in the solar wind, 
we now explore their interactions.

{\it Observations and analysis.}---   Our 
study is motivated by an earlier work that   
examined the  
scalar temperature $T_p=(2T_{\perp p}+T_{\parallel p})/3$ as a function
of $R_p$ and $\beta_{\parallel p}$ and produced evidence suggesting
plasma near the thresholds was hotter than expected \citep{liu06}.
This would be very interesting because the instabilities should not 
heat the plasma, only isotropize it. However, it was 
difficult to tell if the plasma was truly hotter near the 
instability thresholds, or if the result was due to the dependence
of $\beta_{\parallel p}$ on $T_{\parallel p}$.
To better understand these associations,
we use a
more accurate set of temperature measurements and examine
the components $T_{\perp p}$ and $T_{\parallel p}$ separately.

\begin{figure}
  \includegraphics[scale=.50]{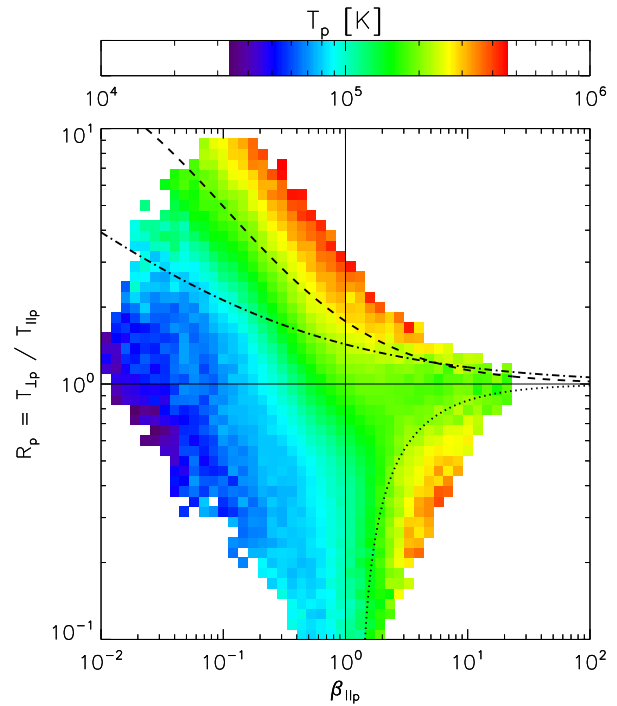}
  \caption{Temperature
    $T_p=(2T_{\perp p}+T_{\parallel p})/3$ as 
    a function of
    $R_p$ and $\beta_{\parallel p}$.  Curves indicate theoretical 
    thresholds due to instabilities.  There is an overall trend of
    increasing $T_p$ with $\beta_{\parallel p}$, which is expected
    since $\beta_{\parallel p}$ is proportional to $T_{\parallel p}$, 
    but there is also a clear association between the thresholds
    and hot plasma.  
    \label{fig2}}
\end{figure}

This study makes use of observations from two 
instruments on the {\it Wind} spacecraft: 
92-second cadence ion velocity spectra 
from the Solar Wind Experiment Faraday 
Cup (FC) instruments \citep{ogilvie95} and 3-second 
measurements of {\bf B} from 
the Magnetic Field Investigation \citep{lepping95}.  
This merged dataset is publicly 
available and has been described in detail elsewhere \cite{kasper06}.  
In all of our earlier work, and to the best of our knowledge in 
all other determinations of anisotropy in the solar wind, higher 
time resolution {\bf B} values were averaged over 
each ion measurement.  $R_p$ was then determined
by examining how $T_p$ depends upon direction relative to the 
average {\bf B}.  We realized that the strong magnetic fluctuations
generated by unstable plasma \citep{bale09} may create 
errors in $R_p$ and rewrote our analysis to use the 
3-second values of {\bf B}.  
This technique produces more accurate and
often larger $R_p$ when there are large fluctuations.  While a detailed
report on this method is in preparation, this letter 
contains our first result.

\begin{figure*}
  \includegraphics[scale=0.5]{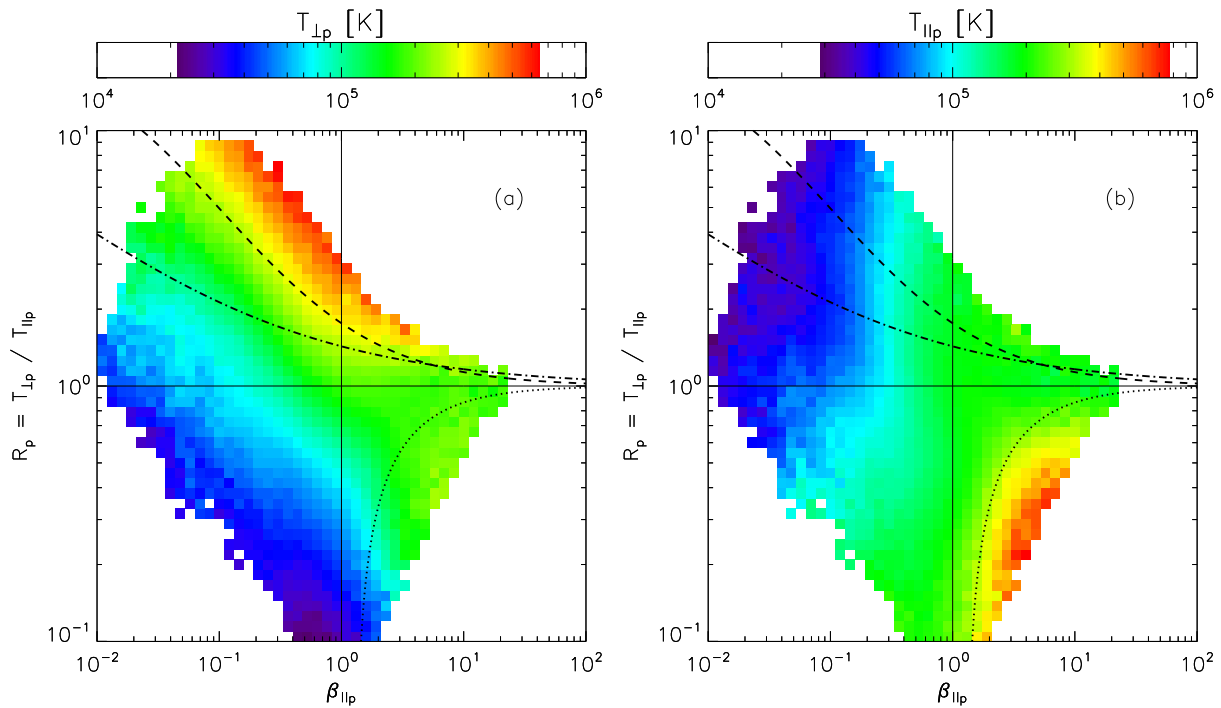}
  \caption{$T_{\perp p}$ (a) and 
    $T_{\parallel p}$ (b) over the $(\beta_{\parallel p},R_p)$-plane.  \
    The increase in $T_p$
    is mainly in $T_{\perp p}$ for plasma beyond the 
    mirror threshold and in $T_{\parallel p}$ for plasma beyond the
    firehose threshold.
    \label{fig3}}
\end{figure*}

About $40\%$ of the $4.1$ million {\it Wind} ion spectra 
met our criteria for use in this study.
We required the uncertainties in the derived temperatures 
to be less than $10\%$.  We only used periods where {\it Wind} was 
far from the Earth's bow shock to avoid magnetospheric 
contamination.  Finally, we only used
observations where Coulomb relaxation was not an 
important factor.  For each measurement, we calculated the Coulomb
collisional age $A_c$ defined  
as the number of small-angle Coulomb scatterings the plasma
has experienced in the time it took to reach the spacecraft 
\citep{kasper08}.  We then required $A_c\le 0.1$.

We divided
the selected observations into a $50\times 50$ grid of logarithmically-spaced 
bins in the $(\beta_{\parallel p},R_p)$-plane.  
Within each bin, we calculated the number of observations $N$ and the 
median value of $T_p$.  For Fig.~\ref{fig1} 
we calculated $f$ by dividing $N$ 
by the width in $\beta_{\parallel p}$ and $R_p$ of each bin.
Fig.~\ref{fig2} shows 
$T_p$ for all $(\beta_{\parallel p},R_p)$-bins with $N\ge 50$.  
Beyond a general tendency for $T_p$ to grow with
$\beta_{\parallel p}$, there are clearly two regions with enhanced
$T_p$: one along the mirror instability threshold and the other 
along that of the firehose instability.
Between these regions, even at 
high $\beta_{\parallel p}$, we see cooler plasma.  
For comparison, $T_p \approx 1.8 \times 10^5 \ \textrm{K}$ 
at $(\beta_{||p},R_p) = (1,1)$ but gets as high as $
T_p \approx 4.6 \times 10^5 \ \textrm{K}$, 
or nearly twice as hot, near the instability thresholds.

Fig.~\ref{fig2} conclusively establishes that there is a 
significant enhancement in $T_p$
near the thresholds for anisotropy-driven instabilities, 
confirming the earlier suggestion of this effect \citep{liu06}.
There is a striking correlation
between the region of increased $T_p$ above the mirror 
and firehose thresholds and the region of 
enhanced magnetic fluctuations
found by \citep{bale09}.  This result further 
confirms the role played by instabilities 
in limiting $R_p$ in the solar wind and the idea that 
the mirror instability is more important than the cyclotron
instability.

We have also calculated  $T_{\perp p}$ and $T_{\parallel p}$
over the $(\beta_{\parallel p},R_p$)-plane.  
The results, 
which are shown in Fig.~\ref{fig3}, are dramatic: the heating near
the mirror threshold is almost 
entirely $\perp$ to {\bf B}, with 
$T_{\perp p}\approx 6.4\times 10^5$ K, 
while the heating near the firehose 
instability is $\parallel$ to {\bf B}, with 
$T_{\parallel p}\approx 7.7\times 10^5$ K.  Close
inspection of Fig.~\ref{fig3} does show an increase, albeit 
smaller, in the other temperature component.    

\begin{figure}
  \includegraphics[scale=0.52]{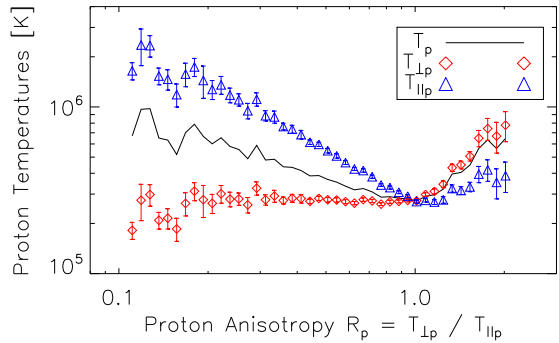}
  \caption{Median scalar proton temperature $T_p$ (black curve)
    and components $T_{\perp p}$ (diamonds) and 
    $T_{\parallel p}$ (triangles) as a function of 
    $R_p$ for all observations with $3\le\beta_{\parallel p}\le 30$.  For
    $T_{\perp p}$ and $T_{\parallel p}$ uncertainty in the mean 
    is indicated with 
    error bars.
    \label{fig4}}
\end{figure}

Consider the high temperature region near the 
mirror threshold.  
In Fig.~\ref{fig4} we plot the median values of $T_{\perp p}$ and
$T_{\parallel p}$ as functions of $R_p$ for all observations with
$3 \le \beta_{\parallel p}\le 30$.  Here we can clearly see
that in addition to an increase in $T_{\perp p}$ for $R_p>1$
there is also a slight increase in $T_{\parallel p}$.
Since we do not expect the mirror instability to 
heat the plasma directly, these results suggest
the following interpretation.  
First, Alfv\'{e}n waves enter cyclotron
resonance with the protons, which increases $T_{\perp p}$ and
thus $R_p$.  Eventually, $R_p$ is sufficiently large that 
the mirror instability sets in and drives the plasma back 
toward isotropy.   In doing so, some of the 
energy deposited into $T_{\perp p}$ by anisotropic dissipation
is transferred to $T_{\parallel p}$.

We now turn to the region near the firehose instability.
Previous work suggests that the 
dominant mechanism producing 
$R_p<1$ is that CGL expansion cools $T_{\perp p}$ more quickly than 
$T_{\parallel p}$, at least
until the plasma reaches the firehose threshold \citep{matteini07}.  
Fig.~\ref{fig4} shows that 
$T_{\perp p}$ is only slightly cooler for $R_p<1$ and
$3\le \beta_{\parallel p}\le 30$, while 
$T_{\parallel p}$ is more than four times 
higher than typical wind, rising to $2$ MK.
This result suggests that a process is increasing $T_{\parallel p}$.
One explanation is that there is an as-yet unknown 
dissipation mechanism in the solar wind  
that can raise $T_{\parallel p}$. 
There is little theoretical work on $\parallel$ heating in
the solar wind but
other recent observational studies have reported cases of unusually 
large $T_{\parallel p}$ \citep{kasper08}.
A second possibility is that 
the heating arises directly as a result of
the plasma being driven into the firehose instability 
by CGL expansion.  Simulations of expanding wind
have suggested that the firehose instability reduces 
$T_{\parallel p}$ but also produces a high energy 
tail of particles $\parallel$ to {\bf B} \citep{matteini06}.  
Perhaps this high energy tail subsequently relaxes and heats the plasma.

The energy for this speculative $\parallel$ 
heating could come from slowing 
the solar wind down or by modifying the expansion of the magnetic field.
The solar wind thermal
energy is typically about $1/100$ the kinetic energy.  So for 
the largest $T_{\parallel p}$ 
about $4\%$ of the 
bulk energy of the wind would have to be converted into thermal energy.  

{\it Conclusions.}--- 
We have shown that protons with anisotropy beyond the
mirror and firehose thresholds are
$3-4$ times hotter than those in 
typical solar wind.  We examined 
$T_{\perp p}$ and $T_{\parallel p}$ and found that
both components are hotter near each instability, but that 
most of the additional heating is in $T_{\perp p}$ for $R_p>1$
and in $T_{\parallel p}$ for $R_P<1$.
This result is interesting because instabilities are 
not understood to heat plasma themselves.  
The hot plasma is therefore either a signature of the interaction of
instabilities with secondary processes or evidence that
instabilities do more than merely redistribute thermal energy.  

For $R_p>1$ we suggest that 
the high temperatures are due to a combination of an ion-cyclotron 
resonant heating process 
(such as dissipation of Alfv\'{e}n waves) increasing 
$T_{\perp p}$ and redistribution of 
thermal energy from $T_{\perp p}$ into $T_{\parallel p}$ by the 
mirror instability.
A plasma undergoing slow $\perp$ heating may achieve 
a state
where the rate of injection of energy into 
$T_{\perp p}$ is balanced by the redistribution of energy from 
$T_{\perp p}$ into $T_{\parallel p}$ by the mirror instability.  Measurements of 
anisotropies alone therefore underestimate the level of heating
from dissipation.

For $R_p<1$ the situation is less clear.  
Either there is an as-yet unidentified parallel heating mechanism 
at work in the solar wind, or the combination of CGL 
expansion and the firehose instability are capable of
converting a small fraction of the bulk kinetic energy of the wind
into thermal energy.

These results have implications beyond the solar wind.  In 
any expanding plasma with $\beta \gtrsim 1$, the CGL-firehose
association may be an effective mechanism for parallel heating
and the generation of magnetic fluctuations.  This
result is directly applicable to heating and magnetic field generation
in clusters of galaxies \citep{schekochihin05}.  
Our results also 
suggest a heating mechanism for contracting astrophysical 
plasmas.  Consider, for example, accretion onto 
a compact object such as a black hole or a neutron star.  There is a great 
deal of kinetic energy gained by the accreated matter as it falls into 
the gravitational potential well of the compact object.
However, it is not understood how this kinetic energy is 
converted into thermal energy in the dynamical time 
of the infall.   We have shown
how expansion of the solar wind drives the plasma into the firehose
instability and ultimately heats it.  In the case of accretion, contraction 
of the plasma would instead drive $R_p>1$, where the mirror 
instability might play an analogous role to the firehose instability in the
solar wind,  heating the accreting material while slowing the 
flow or shearing the magnetic field.

\begin{acknowledgments}
JCK and BAM thank S. Cranmer, J. Raymond, and R. Narayan for 
discussions.  Analysis of Wind observations is supported by 
NASA grant NNX08AW07G.
\end{acknowledgments}

\bibliographystyle{apsrev}

\clearpage


\end{document}